# "TikTok, Do Your Thing":
# User Reactions to Social Surveillance in the Public Sphere

Meira Gilbert
*University of Washington*

Miranda Wei
*University of Washington*

Lindah Kotut
*University of Washington*

## Abstract

"TikTok, Do Your Thing" is a viral trend where users attempt to identify strangers they see in public via information crowd-sourcing. The trend started as early as 2021 and users typically engage with it for romantic purposes (similar to a "Missed Connections" personal advertisement). This practice includes acts of surveillance and identification in the public sphere, although by peers rather than governments or corporations. To understand users' reactions to this trend we conducted a qualitative analysis of 60 TikTok videos and 1,901 user comments. Of the 60 videos reviewed, we find 19 individuals were successfully identified. We also find that while there were comments expressing disapproval (n=310), more than double the number expressed support (n=883). Supportive comments demonstrated genuine interest and empathy, reflecting evolving conceptions of community and algorithmic engagement. On the other hand, disapproving comments highlighted concerns about inappropriate relationships, stalking, consent, and gendered double standards. We discuss these insights in relation to the normalization of interpersonal surveillance, online stalking, and as an evolution of *social surveillance* to offer a new perspective on user perceptions surrounding interpersonal surveillance and identification in the public sphere.

## 1 Introduction

In major cities across the world, the "Missed Connections" personal advertisement section on Craigslist hosts dozens of posts every week [44]. Individuals who post on Missed Connections can successfully identify a stranger they encountered on the bus, at a bar, or walking down the street if the stranger sees their post before it is automatically removed.[1] In other words, the connections that emerge require mutual interest by both the Craigslist post "creator" and the "subject" they are looking for.

On TikTok, the shortform video-based social media platform, a similar "Missed Connections" phenomenon exists but with one key difference: Users post recordings of individuals without their knowledge or consent, leveraging crowd-sourcing in the attempt to identify them.

Recording videos of strangers in public and posting them on the internet is not new nor unique to TikTok. Variations of this trend occur on Twitter and Reddit, and related behavior has recently made the news when two students were able to identify a stranger in public by leveraging smart glasses and facial recognition technology [27].

However, there are significant risks to being recorded and identified by peers, even in public spaces. Surveillance can limit personal freedoms, contribute to chilling effects of free speech, or make individuals feel uncomfortable or uneasy [52]. Identification can similarly impact personal and professional freedoms by infringing upon an individual's ability to be anonymous [52]. Examples of these outcomes might include instances where student protesters are doxed as a retaliation tactic [39], people identified from viral videos of racist outbursts lose employment [54], or ordinary people are made uncomfortable when identified online [16].

To investigate user perceptions of recording someone in public and leveraging social media platforms to identify them, we turn to a viral TikTok trend called "TikTok, Do Your Thing". In the trend, TikTok users ("creators") attempt to identify strangers ("subjects") they see in public. Creators accomplish this by capturing photographs and/or video recordings of the subjects, posting the photographs or videos on TikTok, and asking other TikTok users for help finding and identifying the subject. If successful, the identification generally occurs in the TikTok comments section via information crowd-sourcing.

The virality of this TikTok trend and the high level of engagement in the comment sections offer a unique opportunity to gain insight into user reactions and public discourse on the emergent privacy issues. We conduct a qualitative analysis of 1,901 top-ranked user comments from 60 TikTok videos to make the following two contributions to the field of usable

---
[1]Craiglist ads stay up between 7 and 45 days, depending on locale and type of post [17].

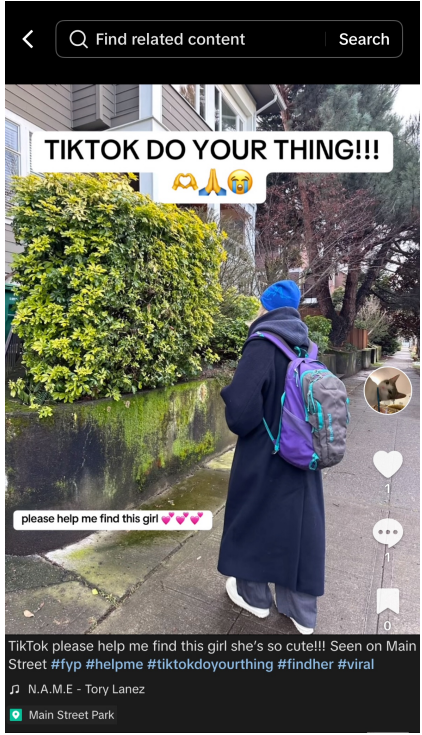

Figure 1: Example of a "TikTok, Do Your Thing" video (recreated to maintain creator anonymity.)

privacy and security:

First, we find that with only a recording, TikTok users identified 19 individuals from these 60 videos. We also characterize user reactions towards this practice, identifying reasons for user (dis)approval. We find that supportive users join the trend's communal identification effort by offering tactical advice or information about the subject, while others leverage algorithmic engagement tactics to boost video views. These users view the trend as a suspenseful, relatable, and innocuous option to genuinely build romantic connection. On the other hand, users who oppose this practice offer negative judgments towards the behavior. These users attempt to discourage the behavior by considering it antisocial or abnormal, offering disapproving relationship advice, and raising concern with perceived stalking, gendered double standards, and lack of consent involved.

Second, we apply theories of social norms and contextual integrity to our findings to explain how a practice with overt potential privacy problems is made "cute" rather than "creepy." We discuss how the "TikTok, Do Your Thing" trend normalizes social surveillance and online stalking. We conclude by discussing the implications and risks of social media-facilitated interpersonal surveillance and directions for future work.

## 2 Background

TikTok, the short form social media video platform, stands out among its peers for its emphasis on video creation and engagement [65]. While it bears many of the standard markers of social media platforms (profiles, friend lists, following options), the platform encourages content generation via acts of imitation and replication, ultimately positioning mimetic behavior as the social basis of the site [65].

The main TikTok feed, dubbed the "For You Page" (FYP), is therefore saturated with similar, imitative videos, and TikTok trends are examples of this imitation in action. On their FYPs, TikTok users are shown public content from unconnected users around the world. TikTok users are able to directly engage with these videos via liking, commenting, sharing (similar to "reposting"), or duetting/stitching (creating new videos linked to the original video content).

We investigated the "TikTok, Do Your Thing" trend in this work. In the trend, TikTok users (who we call "creators") attempt to identify strangers (who we call "subjects") they see in public. Creators accomplish this by 1) capturing photographs and/or video recordings of the subjects, 2) posting the photographs or videos on TikTok, occasionally with additional identifying information, and 3) asking other TikTok users for help "finding" the subject. Creators may also tag or label their video with a location, which allows TikTok to recommend the video to other users located nearby [37]. If successful, the identification generally occurs via information crowd-sourcing in the video comment section. We refer to users who comment on the creator's TikTok video as "commenters."

Although videos within this trend vary in content, style, and intention, users frequently label these videos with some variation of "TikTok, do your thing" (a riff on the internet meme slang, "Internet, do your thing") suggesting a belief that the TikTok trend, platform, or community can successfully help the creator accomplish their goal of finding the subject. Versions of this trend have previously made the news in instances where it was used to identify individuals cheating on their partners [5, 43] and to find a girl in a short TV clip featuring her watching a college football game [58]. For the subject, the trend poses both privacy concerns (potential loss of anonymity and revealment of personal information) and surveillance concerns (recording of their actions, mass monitoring from social media users).

## 3 Related Work

### 3.1 Privacy on Social Media

Prior work on privacy on social media largely focused on how users understand or manage their privacy. For example, usable privacy and security researchers have examined privacy attitudes and concerns on social media [32, 53], as well as privacy-related behaviors on social media, e.g., post sharing [6], information sharing with third-party apps [34], and privacy settings [41]. In contrast, we turn our focus towards content on social media that pose privacy risks to others, and user reactions to that content. Recent work demonstrates how researchers can qualitatively analyze TikTok videos to gain insight into behaviors that pose security risks in interpersonal relationships, e.g., in intimate partner relationships [61] and in parent-child relationships [55, 61]. We further build on this work by demonstrating how TikTok enables accessible, even playful, methods for users to engage in interpersonal surveillance of strangers, guided by collectively defined social norms.

Norms on social media are described by Helen Nissenbaum's theory of privacy as *contextual integrity*, which connects privacy to the norms of the specific contexts it exists in, "demanding that information gathering and dissemination be appropriate to that context and obey the governing norms of distribution within it" [45]. Related work studies how social sanctions - defined as behaviors that encourage or discourage conformity to social norms - enforce and sustain privacy norms on social media [48]. Studies investigating how individually and collectively shared social norms impact user perceptions of online photo sharing [28] and how humor styles impact online photo sharing actions [26] demonstrate the role of norms in impacting user privacy perceptions and behaviors. Other work focusing on privacy in online dating offers information on how user values and goals inform their perceptions and behaviors related to privacy [14]. We build on these works to gain insight into how the "Tiktok, do your thing" trend, and the norms it creates, relates to user perceptions of surveillance and identification in the public sphere.

### 3.2 Social Surveillance and Online Stalking

"TikTok, Do Your Thing" is by nature a surveillance action. Surveillance is a type of information collection that is the watching, listening to, or recording of an individual's activities [52]. This monitoring can occur in public or private, and by the government or others [52]. However, surveillance typically refers to monitoring for purposes that allow nation-states, corporations, or other authoritative figures to manage populations in a hierarchal fashion [40]. There are alternative interpretations of surveillance to describe other dynamics. Examples include *sousveillance*, to define the directionally opposite monitoring of the public toward authorities [38], and participatory surveillance, to define the empowering, playful, mutual monitoring of others on social networking sites [3].

However, the power and social dynamics these definitions present do not sufficiently explain the TikTok trend. The social media facilitated surveillance and identification, as well as the unequal interpersonal dynamics (creator surveilling the subject, other TikTok users surveilling the subject and engaging with the TikTok creator) positions this practice as an act of social surveillance, which Alice Marwick describes as "the ongoing eavesdropping, investigation, gossip and inquiry that constitutes information gathering by people about their peers, made salient by the social digitization normalized by social media" [40].

Social surveillance is distinguished from other types of surveillance by its relationship to power, hierarchy, and reciprocity, and is particularly relevant to explain behaviors on social media, such as monitoring a friend's Instagram posts and knowing they are also monitoring your online activities [40]. Social surveillance differs from traditional surveillance because it occurs between peers and accounts for the power differences evident in everyday interactions rather than hierarchical institutions [40]. In one case study, Marwick uses "Facebook stalking," the practice of using Facebook to discover information about others, to understand how social surveillance is tied to power relations. She argues that Facebook stalking is a way to overcome perceived social weaknesses by gaining social knowledge about others to maintain social hierarchies [40].

The behavior presented in Facebook stalking is now ubiquitous across other social media platforms, where users willingly share personal information across social media platforms [63]. While many subjects in Marwick's study discuss this mode of stalking in a positive manner (e.g., connection with others), researchers have highlighted the potential risks of this practice. For example, Yang et al. examine the vulnerability of private information in online sources, and demonstrate that large portions of users with online presences are identifiable even with limited or inaccurate "seed" information (such as name) [63].

Further work on user perceptions of stalking refers to the social or power differentials involved, such as how gender is a factor in user perceptions of when behavior is considered cyberstalking or illegal [1]. Moreover, information gathered from online stalking creates opportunities for online harassment or abuse, including spamming of social media accounts, doxxing, or potentially even manifesting physical harm such as SWATing [57]. We build on this framing of peer-to-peer, interpersonal social surveillance and the implications of online stalking to address the surveillance dynamics occurring in the "TikTok, Do Your Thing" trend.

### 3.3 Social Media Comment Analysis

Researchers analyze comments on social media with a variety of methods and research goals. This has included content analysis on comments to draw public opinion data and user insights on topics including vaccination [30], sex education [22], COVID-19 [46], body positivity [25, 56] and intimate partner violence [62], as well as dozens of others ranging from food, fashion, and celebrity gossip [2]. Typical comment analysis approaches do so in the context of the news [2]. Specifically, mis/dis-information scholars turn to social media comments to test hypotheses about civility in online political discussion [49], user conformity [15], comment tone's impact on credibility [33], and linguistic signals related to misinformation and fact-checking [31] to improve online discourse.

In their literature review of comment analysis on social media, Alafwan et al. [2] recommend that TikTok be further examined as a source for comment content analysis given the high number of users and low number of quality research on the TikTok platform. Some researchers have begun this work. For example, Armin et al. [4] conduct topic modeling on TikTok comments scraped from videos with the hashtag "#harassment" to understand how public discourse on the topic unfolds on the platform, while Chen et al. [13] conducted text mining on TikTok comments to analyze the rural landscape preferences of video viewers.

We build on prior work by contributing qualitative methods and analysis of TikTok comments.

## 4 Method

To address our research questions and gain insight into user discourse and reactions, we conduct a reflexive thematic analysis of TikTok videos and their associated comments.

### 4.1 Data Collection

Data collection occurred between October 2024 and January 2025. As a TikTok user, the first author had organically encountered relevant videos and was familiar with some common hashtags and phrases used by TikTok creators for the trend. The first author began by searching for videos using TikTok's built-in search interface, iteratively adding search terms as they encountered more videos until no new terms or videos could be found. While keyword searches were the main method used to gather videos, we continued seeking out new videos throughout the data collection process through other methods. For example, we encountered several accounts created for the sole purpose of finding an individual, so we then manually searched for other public accounts with variations of "help me find" in their usernames. We also included relevant videos sent to us by peers aware of our project. This initial exploratory data collection yielded over 130 videos.

Our research questions about the "TikTok, do your thing" trend are about how people react to videos that both depict a stranger in the video (who may not be aware they were recorded) and attempt to identify that stranger. However, not all videos referring to the trend included these key components. Thus, we next applied scoping criteria to filter our initial dataset to videos that included both:

- **surveillance**: watching, listening to, or recording an individual's activities as a type of information collection [52]

- **identification**: attempting link information or data to individuals as a type of information processing [52]

This allowed us to filter our initial dataset to videos with semantically meaningful content for our analysis. For example, videos with identification but without surveillance might include a TikTok user describing someone they met and revealing personal information about them (name, location, occupation, etc.), but not posting any videos or photos of the individual. Videos with surveillance but without identification might include a TikTok user recording an individual in public, but further investigation reveals they already knew each other and the trend was staged.

We further filtered our dataset to focus on videos with similar user motivations for surveilling and identifying strangers. We initially considered videos with many motivations, such as justice or restitution (e.g., identifying a car in a hit and run; identifying someone talking about cheating on their partner), as well as platonic motivations (e.g., attempting to get in touch with a long lost childhood friend; attempting to track down someone captured in a photograph so they can see the image). We ultimately chose to focus on videos where users had romantic motivations because this was the most frequent and prominent motivation we encountered during the data collection process. Similar to a "Missed Connections" personal advertisement, these videos were motivated by a TikTok user's desire to identify a stranger in public who they perceived as attractive.

Our final dataset included 60 videos: 57 in English, two in a mix of English and Filipino, and one in Spanish. The data collection methods used to find the videos in our final dataset is shown in Table 1. We used PykTok, a Python package, to scrape metadata and comments from these videos via the TikTok API. We ran PykTok on Python version 3.12.4, with Playwright version 1.39.0. Access to comments were limited to approximately the top-ranked 30 comments per video (based on user engagement) due to API restrictions. From the 60 videos, we gathered 1,901 comments. Descriptive statistics of the videos are shown in Table 2.

### 4.2 Data Analysis

**TikTok videos.** To analyze video content, we deductively coded key features following a thematic analysis codebook approach [9, 10]. The first and second authors met to discuss features of interest. After developing initial categories together, they each independently reviewed and coded the videos against the codebook, then aligned on any discrepancies.

Identifying instances where the subject was successfully "found" or identified was complex. We determined the identification was successful if a) a commenter who appeared to resemble the subject in the video identified themselves (e.g., "*omg hey that's me!*") or b) several comments tagged the same social media account or wrote the same name repeatedly. We made these determinations based on reviewing not just the comments in our dataset, but also reviewing the replies to the comments and the identities presented by the TikTok accounts involved.

Table 1: We collected data for our study through searches on TikTok using **keywords** as well as through **other** non-keyword methods. Account search refers to searches conducted for throwaway accounts dedicated to finding someone. Search terms with 0 counts are those which were attempted, but did not yield videos meeting our inclusion criteria.

| Method | Count |
|---|---|
| **Keyword searches** | |
| `TikTok do your thing #findhim` | 19 |
| `find this man` | 15 |
| `find this boy` | 12 |
| `Tiktok do your thing` | 8 |
| `find this girl` | 2 |
| `find this woman` | 0 |
| `find this person` | 0 |
| `TikTok do your thing #findher` | 0 |
| `TikTok do your thing #findthem` | 0 |
| **Other** | |
| Shared by other researchers | 3 |
| Account search | 1 |
| Total | 60 |

Table 2: Statistics about our dataset of 60 videos: **mean**, **med**ian, **min**imum, **max**imum, and **standard deviation**. Views are approximate as abbreviated by the TikTok app.

| | mean | med. | min. | max. | SD |
|---|---|---|---|---|---|
| Duration (in s) | 14.09 | 10 | 5 | 173 | 22.02 |
| Likes | 343K | 40K | 38 | 3.6M | 718K |
| Shares | 11.4K | 289.5 | 1 | 518K | 68K |
| Comments | 2.5K | 282 | 2 | 32.4K | 5K |
| Views | 31.5M | 445K | 2565 | 18.4M | 5M |

Lastly, we classified the types of information identifying the subject. We determined what types of information were shared in the videos by reviewing video descriptions, captions, and comment sections, being cautious to only include identifying information that appeared genuine rather than humorous or ill-natured (e.g., "*they go to my school and are weird*" was excluded). We divided this between two categories: identifying information that was shared either by a) the creator of the video, or b) the commenters, if the subject was successfully identified. A summary of these results is shown in Table 3.

**TikTok comments.** To analyze comment content, we inductively coded the data to carry out a reflexive thematic analysis [9, 11], which is intended to be subjective and interpretive [12]. The first and second author, with guidance from the third author, frequently met to discuss appropriate methods for approaching and coding the data. We determined that our data best lent itself to an "artfully interpretive (Big Q, non-positivist, reflexive)" rather than "scientifically descriptive (small q, positivist)" analysis, precisely because we cannot claim to know and describe the intention and fixed meaning of every comment [12, 21]. With this theoretical approach, we view our themes as an interpretation of core ideas or meanings rather than fixed topic summaries [12].

To conduct the reflexive thematic analysis, the first author reviewed all comments in the dataset to understand the breadth of the content, taking notes and generating memos with initial thoughts and observations. Next, they selected a random subset of the data (190 comments, or 10%) and conducted inductive open coding to determine initial categories and understand what types of content existed in the dataset. They then applied the coding to the remaining data, working to modify, expand, or condense codes as needed.

After several rounds of open coding and review, the first and second authors reviewed the codebook. The second author then independently reviewed the initial codes for the entire dataset, documenting disagreements, questions, and potential changes. After both authors had reviewed the entire dataset, they met again to revisit the codes, align conflicting codes, and finalize the codebook. After completion of the codebook, the two authors conducted further discussions to generate themes from the distinct codes [9]. A codebook organized by themes and code is included in the Appendix (Table 4).

### 4.3 Ethical Considerations

Studying user generated content from social media, even when that data is public, poses potential harms and ethical considerations [64]. Although there is limited existing work on the ethics of collecting user-generated data from TikTok, previous research establishes that users across Twitter, Reddit, and Facebook demonstrate levels of discomfort with this practice, largely based on the context of the research [18, 19, 24].

We turn to existing ethical work in this field urging researchers to "remember the human" when conducting studies leveraging public user generated content [20]. We took several steps to maintain the anonymity and privacy of the TikTok creators, commenters, and subjects involved in our dataset. First, we give only high-level descriptions of the videos without identifiable information such as location or date posted. Second, we do not share the names or TikTok account handles of any user referenced. Third, even though the video content is publicly available, including screenshots in research is ethically fraught [50]. We chose to exclude screenshots, instead opting to re-create similar visuals when needed. Fourth and lastly, we paraphrase all comments to the best of our ability while maintaining the integrity of the comment. While we acknowledge these solutions are imperfect, we believe these steps generate the optimal level of detail required for our analysis while maintaining respect and privacy of the creators and commenters.

Table 3: Of our dataset of 60 TikTok videos, we summarize the **type of account** (a user's personal account or an account dedicated to finding someone), the **type of recording** of the subject, whether the **subject** appeared to be **aware** they were being recorded, whether the subject was **successfully identified**, the **video setting**, who the subject was **identified by**, and the types of information revealed about the subject **in the video** and **in the comments** to the video.

| Account Type | | Recording Type | | Subject Aware? | | Successfully Identified? | |
|---|---|---|---|---|---|---|---|
| Personal | 56 | Video | 49 | No | 40 | No | 38 |
| Find Them | 4 | Photo | 7 | Yes | 20 | Yes | 19 |
| | | Video + Photo | 4 | | | Unclear | 3 |
| **Video Setting** | | **Info Revealed in Video** | | **Identified by?** | | **Info Revealed in Comments** | |
| Concert or music festival | 12 | Face - *full* | 40 | Commenters | 10 | Social media handle | 13 |
| Pedestrian area | 8 | Face - *partial* | 15 | Creator | 7 | Full name | 7 |
| Highway or road | 7 | Place of employment | 11 | Subject | 2 | Nationality | 1 |
| Airplane or airport | 7 | Back of the head | 10 | | | Age | 1 |
| Sporting event | 7 | Vehicle | 6 | | | None | 3 |
| Mall | 4 | Jersey number | 2 | | | | |
| Hotel or resort | 4 | Family information | 1 | | | | |
| Party or nightlife | 3 | Other | 1 | | | | |
| Restaurant | 3 | | | | | | |
| Other | 3 | | | | | | |
| Amusement park | 2 | | | | | | |

### 4.4 Limitations

Manually reviewing 60 videos and 1,901 comments presents several limitations, the key of which is scale. For this reason, the claims of this work are best evaluated as a deep analysis of a specific subset of content, rather broad quantitative claims about an entire online ecosystem.

We were only able to analyze approximately the top-ranked 30 comments per video, where ranking was determined by the TikTok algorithm. However, top comments likely represent those with the most engagement through likes and replies. Comments that video creators engage with (liking, replying, or "pinning" the comment) also appeared highly ranked.

Additionally, it is possible that users who encounter one "TikTok, Do Your Thing" video encounter additional similar videos, or videos from the trend are shown to one specific user group more frequently than another. Therefore, this analysis may be more representative of a specific subset of TikTok users rather than the platform's population more generally.

Similarly, the nature of the data analysis stripped from its original context provides insight towards user reactions in situ, but privileges reactions of users who actively engage and comment on videos. Our analysis does not represent viewers who encounter this trend but do not write a comment expressing that reaction. Nevertheless, visible comments likely shape the opinions of viewers who do not comment, thus meriting our close study.

Lastly, our selection criteria is also limited by our understandings of the videos and does not include any demographic information about the parties involved (creators, subjects, and commenters). Consequently, we are unable to investigate how demographics might impact the privacy attitudes of TikTok users and how those attitudes contextualize our findings.

## 5 Findings

### 5.1 Video Characteristics

Our scoped dataset included 60 videos from the "TikTok, Do Your Thing" trend related to romantic motivations. Similarly to a "Missed Connections" post on Craigslist, creators posted these videos to find an attractive stranger (the "subject") they fleetingly encountered or saw in public. In our videos, the amount of direct contact between creators and subjects was minimal, and the majority (40/60) indicated the subject was not aware of being recorded at all (Table 3).

These videos typically used words and/or emojis in the captions or descriptions to convey a playful romantic interest in the subject, such as subtle expressions of attraction (*"Tik tok do your thing and find this man 🤞 😍"*), mentions of "crushes" (*"airport crush ❤️"*), or hyperbolic statements about love (*"Found the love of my life in [city name] 💍 👰 ⛪ marry me sir."*). While several videos appeared to have specific romantic outcomes in mind, such as getting the subject's phone number (*"[...]My highway crush for 15+ minutes i need his name and phone number ASAP[...]"*) or setting the subject up with someone (*"[...]help me find this man for my friend, they might be a match![...]"*), the vast majority demonstrate playful, casual expressions of romance or attraction without a specific intention or goal.

Another defining trait of these videos is the frequent use of a clip from the song "N.A.M.E" by Tory Lanez as the soundtrack, an apt choice given the song describes falling in love with someone who does not know your name:

> *"(And I swear it, baby) / I fell in love with somebody, who doesn't even know my name / Know my name, no, no"* [36]

Of the 60 videos in our dataset, 19 subjects were successfully identified: 10 were identified by other TikTok commenters, seven were identified by creators (who gave updates in the comment sections), and two subjects identified themselves (Table 3). We observed that most creators posted these videos from their personal accounts but found four accounts specifically created for the purposes of identifying the subject. These accounts had handles such as "@helpmefindthisperson.12345" and posted no additional content beyond the videos in our dataset. However, only one of the 19 successful identifications referenced above were found by "find them" accounts.

The recordings generally occurred in public spaces, and the most frequent setting was cultural events such as concerts or music festivals. Other highly trafficked locations, such as pedestrian areas, airports, malls, resorts, and amusement parks were also common contexts. Interestingly, seven videos were captured "on the road," where the urgency of capturing the subject is most apparent.

Aside from subjects' faces, little personally identifiable information was captured in the videos. However, if identified, commenters would share information including the subject's social media handle, full name, nationality, and/or age. Of the 19 subjects identified, only three videos revealed no further information about them in the comments.

### 5.2 User Responses

Of the 1,901 comments initially analyzed, 691 (36.35%) did not contain overt user reactions. These comments included indecipherable filler language ("*OMG*") spam ("*CLICK HERE FOR 1000*!!!"), nondescript emojis ("⭐"), and tags of other TikTok social media accounts ("@*foo check this out*"). These comments also included jokes, observations, or questions about the videos unrelated to the trend ("*where's your shirt from?*"). While users may have written these comments to express a strong sentiment or reaction, without additional information or context it is was impossible for us to understand the motivation behind them or relate them to our research interests.

The remaining analysis concerns the remaining 1,210 comments (63.65% of our total dataset). We group these comments into two meta-themes (supportive comments and disapproving comments) and 10 sub-themes. The majority (883, or 46.5%) of top comments aligned with generally supportive themes while the minority (310, or 16.3%) were disapproving themes. We counted 17 comments (0.9%) that included codes from both supporting and disapproving meta-themes. Because some comments possess more than one code, our total count of cumulative codes within each sub-theme (1,301) is higher than unique remaining comments in our dataset (1,210). Sub-themes are presented in order of descending frequency.

### 5.3 Supportive Comments

In our reflexive thematic analysis, we interpreted comments expressing support towards the videos as including themes of community building, suspense, relatability, and algorithmic engagement. Our findings suggest that users positively engaged with these videos because they perceive the task of finding the subject romantic and the trend itself suspenseful and engaging, indicating both support and shared responsibility.

**Search Party (n=235).** We found a significant number of comments related to identifying the subject. Within this theme, the majority (n=186) of commenters disclosed they knew the subject or indicated they were someone close to them, such as a friend or family member ("*that's my friend they go to my school*"). While we acknowledge many of these identifications may be overtly humorous ("*I know him he's my brothers sisters cousins best friends dog*"), we classify them as generally supportive, as they reinforce the goal of finding the subject.

We also found instances (n=49) of commenters giving the creators advice on how to find the subject. This advice fell into three major categories: clarifying questions, ("*girl whats the name of the hotel you were in!!!*"), encouraging advice ("*Just go up say I like your vibe and introduce yourself*") and identification advice. Identification advice mainly included strategies to leverage publicly available information in combination with the information obtained from video. For example, if the subject was recorded wearing a uniform

(e.g., police) or in a vehicle with identification numbers (e.g., a firetruck), commenters would suggest contacting the local offices to ask about the subject. Similar advice was given for subjects in sporting gear with jersey numbers, as team websites publicly post their players and numbers, allowing full identification of the player. If the subject was recorded at their job (such as a bar or restaurant), commenters also encouraged the creators to call their place of work and ask about them.

Some tactical advice suggested leveraging third party tools or individuals to help identify the subject. These included apps to make images less blurry or references to professional "profilers" ("*Is there a professional profiler that can figure out where he is based on things in his room?*").[2] Other users suggested conducting a reverse-lookup of the location the subjects were found in to gather more information about the subject:

> "*Did you search the bar as tagged location on all socials? I once found a cheater from a tagged photo at a concert... maybe he's tagged there and you might get even clearer pictures*"

The comment refers to a feature on major social media platforms (TikTok, Instagram, Facebook, SnapChat, and others) that supports location search, where users can see all publicly tagged content from any given location.

**Communal Support (n=217).** Supportive comments also included explicit words of encouragement and support ("*that's so cute, good luck finding her!*"), often using inclusive "we" ("*Damn we all wanna know who he is* 😍😍") and acknowledging a communal role in the search ("*Okay but if we help you find her we can all come to the wedding right?*"). Creators mimic this communal language when relaying updates ("*GUYS WE FOUND HIM!!*").

Commenters also called out the role of TikTok as a platform in hosting this trend, with several comments including variations of "*TikTok do your thing*," "*TikTok do your magic*" and "*the power of TikTok*." One comment highlights how TikTok succeeds at fostering this community compared to other platforms, writing "*I* ❤️ *TikTok - you ask this question on other platforms and people think you're crazy! Find him girl!!!* 🙌".

**Algorithmic Engagement (n=175).** Prior research establishes how TikTok users believe liking, commenting, sharing, and other engagement tactics can manipulate TikTok's algorithms [35]. In the context of this trend, we found that users demonstrate support by commenting to help the creator find the subject. By writing comments to "boost" the videos and referencing their own locations, users believe the videos are more likely to gain views and reach the location where the subject, or people who know them, are most likely to be located.

These comments either reference the location of the commenter ("*Massachusetts here so you're definitely getting a good view of New England* 🤣"), explicitly boost the video ("*I don't know this guy but commenting to boost this video*"), or both ("*we need to boost this so she sees this & you both get your happy ever after* 🥺 *where is everyone from?* 💗").

**Suspense (n=166).** Commenters also indicate support by requesting updates on the search, affirming their investment, or both ("*I'm invested I need an update when y'all are talking and going on a date* 😂"). These comments may indicate a strong user desire to be updated on the search, demonstrating how the task at hand (finding and identifying the subject) is suspenseful and engaging.

**Relatability (n=166).** The last category of supportive comments demonstrates how commenters relate to the creator's desire to find the subject. Comments included those we coded as "empathy" (n=98), where users claim they would do the same as the creator because it is cute ("🥰 🥰 *This is so cute haha. I love it and would definitely do something like this* 😄") or easier than approaching the subject in public ("*This is so me.. I'm way too awkward to just walk up to someone and start talking.* 😭"). Other comments (n=49) reference the attractiveness of the subject ("*He's so hot* 🔥"), suggesting they understand the desire to build a connection. In these various ways, by expressing that they relate to the creator of the video, commenters are implicitly demonstrating approval for the act of identifying the subject.

## 5.4 Disapproving Comments

In our reflexive thematic analysis, we interpreted disapproving comments through the following themes, highlighting the relevance of social norms and demonstrating user concern regarding inappropriate relationships, stalking, gendered double standards, and lack of consent.

**(Anti-)social Norms (n=144).** Majority of oppositional comments reference perceived divergence from established social norms. Versions of the phrase "*[they are] right there*" and "*just talk to [them]*" occur in our dataset 74 and 52 times, respectively. These commenters are unsympathetic to creators who chose to record a video of the subject and post it to TikTok rather than interact with the subject in person ("*so you didn't get a chance to talk to him but u got a chance to record him?*").

When not outwardly disparaging the creator, these comments are sarcastic and demonstrate an unwillingness to help the creator identify the subject. For example, in response to creators asking other TikTok users for help, commenters would write *"You found her already 😂"* or *"you're the closest one to them right now lol,"* demonstrating how users, contrary to those engaged in the participatory, communal aspects of this trend (Section 5.3, **Search Party and Communal Support**) believe the act of finding the subject is the responsibility of the creator and are unwilling to play along. As one snarky commenter writes, *"I ain't google 🙄."*

---
[2]This may be a reference to GeoGuessrs, highly skilled players of an geography game who can quickly identify a location worldwide with very few visual clues [23].

Additional comments within this theme more overtly reference notions of social norms (n=18) by opposing other comments that encourage the trend, (*"Let's normalize going up to people"*) suggesting the videos are abnormal (*"Why don't u just go up to him and talk to him like a normal person instead of taking a video of him and asking tik tok"*), or discouraging the use of TikTok / the Internet to form relationships(*"Idk [I don't know] why people don't interact with each other and have the internet find everyone for them, its lame to say y'all met online"*).

**Relationship (n=89).** Other disapproving comments are related to the appropriateness of the relationship between the creator and the subject. Within this theme, 46 comments referenced the perceived relationship status of the subject, ranging from observations (*"he has a ring on"*) to direct attempts to discourage the creator (*"I hope he's married. People be posting people on here asking without knowing."*). Commenters also related to the subject's potential partner, and how they might feel weird if they saw these types of videos made about their partner (*"imagine his girlfriend when she sees this 👁👄👁"*). In these comments, users appeared to judge the videos more as disrespectful to the theoretical partners rather than focusing on the privacy of the subject.

Some comments (n=25) referenced the perceived ages of the subject and creator. Users expressed concern both if the subject appeared very young, e.g., *"he looks a bit young right...? 💀"* or very old, e.g., *"goodness. TRY TO FIND SOMEONE YOUR OWN AGE!!."*

Other comments (n=18) relied on other justifications to give the creator relationship advice, either dismissing the creator's desire to find the subject based on lack of uniqueness (*"honey there's thousands of carbon copies of the same boy everywhere"*), or referencing personal experiences to discourage the creator (*"From personal experience... run girl"*).

These comments reflected user interest in details of the potential romance rather than any general social privacy harms.

**Fear Factor (n=76).** Within this theme, we found 40 comments suggesting that users had negative emotional judgments or reactions to the videos. These comments leveraged language around fear (*"this comment section makes me fear for my life 💀,"* *"Y'ALL IMAGINE HOW SCARED THE GUY WILL BE!"*), concern (*"It's concerning how you all know so much about him,"*), or disgust over creator's behavior, using words like "cringe" or "weird" as descriptors. Commenters within this theme occasionally put themselves into the subject's shoes, asking what would happen after the subject was found (*"now what are you going to do with him lol"*), feeling scared on their behalf (*"honestly, if I was him I'd be scared"*) or explicitly suggesting that the act of identifying him would negatively impact the subject's life: *"As soon as someone tells u who he is, is as soon as that guy's life is ruined."*

We also found comments (n = 26) explicitly leveraging language surrounding stalking (*"isn't that stalking?,"* *"Stalker-core,"* *"did u really just stalk him..."*). Others (n = 10) reference institutions or people associated with surveillance such as the Federal Bureau of Investigation (FBI) (*"I SWEAR TO GOD TIKTOK COMMENTS ARE BETTER THAN THE F.B.I 😂"*), detectives (*"🕵️🕵️🕵️🕵️🕵️🕵️🕵️🕵️🕵️"*), and the police (*"honestly y'all better than the cops, go find missing ppl haha"*). To these commenters, the videos are creepy, and the creators akin to stalkers.

**Gender (n=17).** While the gender breakdown of the creators and subjects is indeterminable, we observe that in the majority of cases, commenters perceived the creators as women and subjects as men. Subsequently, users leveraged a double standard argument to oppose this practice, claiming the trend was only well received because of the gender dynamics at play. For example, one commenter wrote *"and if the genders were reversed..."* and another *"if a man did this, they're going to jail."*

While we were unable to test this claim given the lack of metadata about creator and subject gender identities, it is interesting to observe that users *assume* this double standard exists, even when there are several viral, well received videos that might suggest otherwise.

**Consent (n=16).** A small number of comments reference the lack of consent (*"please don't take videos of ppl without their consent,"*), permission (*"Why are you taking images of random people without their permission in public 😐"*), or respect (*"I guess you don't know the word respect"*). Within this theme, two comments reference the potential illegality of this recording behavior (*" This is illegal 😐 you don't film someone without their permission 😒"*). The only comment in our dataset that mentioned the word "privacy" also occurred in this theme: *"Miss creator respect their privacy [...]"* (translated from Filipino).

## 6 Discussion

This trend demonstrates how with only a video recording, TikTok users can effectively identify individuals in public spaces via information crowd-sourcing. Across 1,901 user comments, we find a rich set of user reactions to this practice. We relate our findings to contextual integrity and the normalization of surveillance, online stalking, and social surveillance practices. To add further context to our qualitative analysis, we include additional commentary on selected replies to comments in our dataset that were not part of our original analysis. However, this supplementary material is not intended for claims about our broader dataset.

## 6.1 Normalization of Surveillance

In sociology, social norms may be defined as implicit, unspoken rules or behaviors that are not formally written or recorded but are understood by a social group [48]. To be upheld, norms are typically accompanied by social sanctions, which are the penalties and rewards that reinforce conforming behavior [48].

Our analytical interpretations in this work offer insight into how a practice that possesses potential privacy harms is "normalized" within the "TikTok, Do Your Thing" trend. We find that supportive comments both define and perpetuate the "rewards" for participating in the trend. As we discuss in Section 5.3, supportive commenters communally support the identification efforts, offer words of encouragement, affirmation, and advice, relate to the behaviors exhibited in the trend, and algorithmically engage with these videos to generate more video views. To the video creators, these positive "rewards" validate, encourage, and facilitate their efforts to identify the subject. The videos garner high engagement, supportive comments, and TikTok virality or fame.[3] TikTok users are encouraged to participate in acts of imitation and repetition [65], and are therefore motivated to conform to norms of this practice. The result of these positive sanctions is an established social norm defining this trend as cute, romantic, and innocuous.

Applying Helen Nissenbaum's theory contextual integrity explains how the social norms created by and within the trend mitigate its potential privacy concerns. The theory explains that information disclosure is related to the norms of the contexts it exists in [45]. Therefore, if the behavior exhibited in the TikTok trend is an established "social norm," and creators who post these videos follow the norms of the trend, users have valid reasons to lack concern with the practice: The act of recording and identifying a stranger, within this context, is made appropriate.

The justifications offered by commenters align with contextual integrity, though they do not use that exact term. For example, in response to a comment asking "*why is nobody hating,*" a user claims that the context the behavior exists in (a "common" type of TikTok video) and the scope of information revealed (back of head and side profile only) is different than nefarious acts such as doxing:

---
[3]Also called "online clout."

*"Asking tik tok to find people is common, she didn't dox him or even show his face... it was the back of his head and a side profile and you're all acting like she showed his home address"*

Contextual integrity also explains why some disapproving comments counteract the norms of the trend (Section 5.4, **(Anti)-social Norms**). By attempting to classify behavior presented in the "TikTok, Do Your Thing" trend as diverging from established social norms, these comments attempt to remove or limit the context that normalizes such behavior. While privacy concerns can be held as social norms, we also interpret these comments as an indication these users reacted negatively to the trend due to perceived violation of social norms rather than solely privacy concerns, as we would have initially predicted.

Attempts to identify someone for romantic purposes is not novel or unique to TikTok; the regional "Missed Connections" listings popular on Craigslist now exist as subreddits [47] or standalone websites [42]. Additionally, some of the video characteristics we observed in Section 5.1, such as limited or fleeting interaction between creator and subject and an emphasis on physical attractiveness, mirror the content of "Missed Connections posts [8].

However, this trend builds on "Missed Connections" by including acts of surveillance (i.e., video recording), and leveraging the TikTok algorithm and information crowd-sourcing to identify a subject without their knowledge. Previous analysis of TikTok trends and the privacy norms they reinforce or establish demonstrate how TikTok enables accessible, even playful, methods for users to engage in surveillance in multiple contexts [55, 61]. We similarly conclude that the "TikTok, Do Your Thing" trend, and the norms it establishes, creates additional opportunities for users to move beyond the behaviors exhibited by "Missed Connections" to engage in interpersonal surveillance, normalized, facilitated, and made accessible on the TikTok platform.

Overall, even if some commenters appear unconcerned about the privacy implications of this trend, our work demonstrates how the popularity of the trend may risk normalizing interpersonal surveillance [51]. When the trend is accessible, playful, and generally well received, it makes the act of being watched, observed, and identified more palatable and carefree. This builds on existing work from surveillance studies describing "squeeveillance," or "the use of cuteness to normalize surveillance," [7] and exacerbates findings on constant vigilance and user fatigue [51]; eroding the "right to be left alone" [60].

As we discuss in the following two sections, this erosion of privacy and normalization of surveillance poses significant risks for user safety both online and offline.

## 6.2 Online Romance, or Online Stalking?

With only an image, TikTok users can find and identify subjects by name and social media handle (Table 3). Though the behavior is motivated by romantic intent, publicly disclosing this information to millions of TikTok users creates opportunities for online harassment or abuse of the subjects, including spamming their social media accounts, doxing them, or potentially even manifesting physical harm such as SWATing [57]. Even if the harms are not physical, identification online can lead to discomfort and unease from those involved [16].

While most disapproving comments challenge the social norms of this practice, some directly articulate the potential privacy risks and harms present in these videos. We found comments referencing stalking (Section 5.4, **Fear Factor**) particularly interesting, as they suggest users may understand this trend as an example of online stalking or other forms of harassment.

The risks of this practice make user reactions conveying concern and fear, as well as comparisons to stalking, understandable. Viewing the behavior as online stalking also explains the perceived gender double standard involved (Section 5.4, **Gender**). The opinion aligns with previous research demonstrating perpetrator gender as a significant factor in user perceptions concerning cyber stalking [1]. Specifically, people may be more likely to view a given scenario as an example of online stalking and being illegal when the perpetrator is a man compared to a woman [1].

To counter narratives about stalking, some comment replies noted how a video is not stalking due to the public nature of the act:

> *"are you that chronically online that you think someone recording someone at a public event and thinking he's cute is the same as stalking?"*

Other user discourse alternated between those advocating for respecting people's privacy and asking for consent, and those asserting that there is no right to privacy in public. The following abbreviated exchange between three users in the comment replies highlights this argument:

> User A: "*dude she literally recorded him without this consent even without context thats bad have u guys ever heard of respecting ppls privacy*"
> User B: "*its a public place bro.*"
> User C: "*You telling me since I decided to leave the house I now give the entire world permission to record me?...*"
> User B: "*Yes.*"

Notably, U.S. courts agree with User B and hold that there is little to no reasonable expectation of privacy in public spaces [52].

In the absence of legal regulation, subjects have little recourse if they are uncomfortable being featured in the videos or believe they are being stalked. Their only options occur within the TikTok app: Contacting the creator and asking them to remove the video, or reporting the video to TikTok and hope it gets removed. TikTok's Community Guidelines do prohibit sharing content that "*includes personal information that may pose a risk of stalking, violence, phishing, fraud, identity theft, or financial exploitation. This includes content that someone has posted themselves or that they consented to being shared by others*" [59]. However, previous incidents suggest TikTok does not consider these types of videos a violation [16]. Most importantly, subjects are likely unaware they are featured in these videos at all, especially in cases where they are not TikTok users and / or the video does not gain significant viewership.

Building on user comments expressing concern regarding a (lack of) consent (Section 5.4, **Consent**), we believe that consent between creators and subjects would promote user agency over their likeness and identity, as well as mitigate the above concerns regarding stalking. However, this particular trend would likely not occur if the creator gained consent from the subject. Unlike other forms of online stalking where attackers leverage "seed" information users willingly post online [63], the subjects in the "TikTok, Do Your Thing" trend did not willingly post information about themselves for others to access or consent to the behavior in advance.

As a potential solution, we propose building on Jane Im et al.'s [29] work on affirmative consent, where they call for the design of sociotechnical systems that better leverage principles of consent for online engagement. Specifically, we are interested in exploring what technical or design tools could better promote consent and allow more agency over identity and anonymity on online platforms. We strongly consider future work evaluating user interest or adoption in these systems, their risk perception of online identification practices, and the efficacy and safety of such tools.

## 6.3 Evolution of "Social Surveillance"

In addition to interpreting this behavior via social norms and as a form of online stalking, we raise an additional lens through which to understand and analyze this practice: social surveillance [40]. We propose extending the term to describe surveillance occurring between peers not just *on* social media, but facilitated *by* social media.

Our analysis of the "TikTok, Do Your Thing" videos and comments highlight the virality of the trend, the communal effort required to "boost" these videos, and the identification advice and information users are willing to share to support the search. Videos in our dataset gained millions of views and facilitated successful identification of 19 individuals. Supportive comments demonstrates how many TikTok users are comfortable "finding" others, together.

However, subjects of the "TikTok, Do Your Thing" trend do not know they are being monitored by others and have not consented to providing their information on social media platforms. The social media facilitated surveillance and identification, as well as the interpersonal dynamics at place (creator surveilling the subject, and other TikTok users monitoring and engaging with the TikTok creator) positions this practice as an act of social surveillance.

Therefore, the behaviors in these videos pose not just privacy concerns (revealing identifying information, lack of anonymity) but also surveillance concerns. Even though social surveillance occurs between peers, excessive monitoring of others may cause users to modify their own behavior accordingly or reinforce existing power dynamics between individuals [40]. Instances of identifying strangers via social media, for a wide variety of motivations, severely limits the personal freedoms of those involved [52] by bringing unwanted attention and / or enabling doxing [5, 16, 39, 43, 54].

Additionally, the social surveillance behavior exhibited by trend is not an anomaly. Recent consumer based identification tools such as PimEyes are actively used to identify ordinary individuals for social purposes [27], and there are TikTok accounts leveraging these tools to locate and identify individuals for user amusement, highlighting the popularity of open source intelligence (OSINT) [16].

The ease of TikTok and other social media platforms to facilitate this behavior highlights the need for further research unto potential risks and mitigation of interpersonal surveillance, especially those facilitated on social media platforms by other users. Whether occurring via information crowdsourcing, facial recognition technology, or other tools, we urge researchers to investigate the risks and harms these practices pose.

### 6.4 Future Work

Future work could explore user reactions towards other types of surveillance and identification content on TikTok, such as videos for restitution or platonic motivations. Additionally, this work could leverage NLP to gather more videos and comments from a wider range of videos, and use quantitative analysis methods to analyze user discourse at scale. While we included some comment replies as additional context in our discussion, future work is also needed to gather replies and nested comments, which appear to be a rich source of user discourse. As these comments are more argumentative and rooted in debate, further analysis of comment replies would likely reinforce the existing themes identified, but could also demonstrate new areas for inquiry.

TikTok comments offer insight into direct user reactions unmediated by research tools such as survey instruments or interviews, but are limited to users who actively engage and comment on videos. Additional work is needed to explore the perceptions of non-active users who "lurk" (only view content without commenting) or non-TikTok users entirely. We are also interested in investigating how commenter demographics and TikTok's algorithmic targeting could impact observed user reactions towards this trend. Research in these areas will provide more insight into how a variety of user behaviors, user demographics, and platform features impact privacy attitudes on TikTok.

Lastly, researchers interested in assessing this phenomenon longitudinally could conduct an evaluation of shifting interpersonal privacy norms and platform affordances (e.g., Craigslist vs. TikTok). Additionally, while our analysis did not surface meaningful differences in the behaviors comments were critiquing (e.g., making videos vs. sharing vs. sharing broadly), we believe further analysis of disapproving user comments could provide additional nuance into the

Ultimately, a thorough understanding of user perceptions is urgently needed to inform the design of social media platforms, online tools, or physical devices (such as smart glasses) that enable surveillance and identification to occur in the public sphere.

### 7 Conclusion

Analyzing a TikTok trend where users record people in public and attempt to identify them online offers insight into user reactions towards this practice. We compare and contrast why users both support and oppose this practice, finding that supportive users demonstrate genuine interest and empathy, reflecting the communal and supportive nature of the trend. On the other hand, disapproving comments highlight the (anti)social aspects of the behavior and user concerns with inappropriate relationships, stalking, gendered double standards, and consent. Applying theories of social norms and contextual integrity to our findings demonstrate how and why this practice is both normalized and challenged by users despite potential privacy harms. Additionally, our findings shed light into how this behavior relates to user perceptions of online stalking, and builds on prior work investigating social surveillance practices. We conclude with calls for increased attention towards interpersonal social surveillance, especially those occurring on and facilitated by social media platforms.

### Acknowledgments

We thank the anonymous SOUPS reviewers for their constructive feedback. We also thank Dr. Tanu Mitra and Stephen Prochaska for early feedback guiding research design and methodology.

## References

[1] Billea Ahlgrim and Cheryl Terrance. "Perceptions of Cyberstalking: Impact of Perpetrator Gender and Cyberstalker/Victim Relationship". In: *Journal of Interpersonal Violence* 36.7 (Apr. 1, 2021). Publisher: SAGE Publications Inc, NP4074–NP4093. ISSN: 0886-2605. DOI: 10.1177/0886260518784590. URL: https://doi.org/10.1177/0886260518784590.

[2] Brian Alafwan, Manahan Siallagan, and Utomo Sarjono Putro. "Comments Analysis on Social Media: A Review". In: *ICST Transactions on Scalable Information Systems* (Sept. 6, 2023). ISSN: 2032-9407. DOI: 10.4108/eetsis.3843. URL: https://publications.eai.eu/index.php/sis/article/view/3843.

[3] Anders Albrechtslund. "Online social networking as participatory surveillance". In: *First Monday* (2008). ISSN: 1396-0466. DOI: 10.5210/fm.v13i3.2142. URL: https://firstmonday.org/ojs/index.php/fm/article/view/2142.

[4] Atieh Armin, Joseph J Trybala, Jordyn Young, and Afsaneh Razi. "Support in Short Form: Investigating TikTok Comments on Videos with #Harassment". In: *Extended Abstracts of the CHI Conference on Human Factors in Computing Systems*. CHI EA '24. New York, NY, USA: Association for Computing Machinery, May 11, 2024, pp. 1–8. ISBN: 9798400703317. DOI: 10.1145/3613905.3650849. URL: https://dl.acm.org/doi/10.1145/3613905.3650849.

[5] Beth Ashley. *My Relationship Was Derailed By A Viral TikTok*. URL: https://www.refinery29.com/en-gb/2021/09/10615723/viral-tiktok-cheating-boyfriend.

[6] Oshrat Ayalon and Eran Toch. "Retrospective privacy: managing longitudinal privacy in online social networks". In: *Proceedings of the Ninth Symposium on Usable Privacy and Security*. SOUPS '13: Symposium On Usable Privacy and Security. Newcastle United Kingdom: ACM, July 24, 2013, pp. 1–13. ISBN: 978-1-4503-2319-2. DOI: 10.1145/2501604.2501608. URL: https://dl.acm.org/doi/10.1145/2501604.2501608.

[7] Garfield Benjamin. "Squeeveillance: Performing Cuteness to Normalise Surveillance Power". In: *Surveillance & Society* 22.4 (Dec. 6, 2024), pp. 350–363. ISSN: 1477-7487. DOI: 10.24908/ss.v22i4.16673. URL: https://ojs.library.queensu.ca/index.php/surveillance-and-society/article/view/16673.

[8] Jennifer L. Bevan, Jimena Galvan, Justin Villasenor, and Joanna Henkin. ""You've been on my mind ever since": A content analysis of expressions of interpersonal attraction in Craigslist.org's Missed Connections posts". In: *Computers in Human Behavior* 54 (Jan. 2016), pp. 18–24. ISSN: 07475632. DOI: 10.1016/j.chb.2015.07.050. URL: https://linkinghub.elsevier.com/retrieve/pii/S0747563215300601.

[9] Virginia Braun and Victoria Clarke. "Using thematic analysis in psychology". In: *Qualitative Research in Psychology* 3.2 (Jan. 2006), pp. 77–101. ISSN: 1478-0887, 1478-0895. DOI: 10.1191/1478088706qp063oa. URL: http://www.tandfonline.com/doi/abs/10.1191/1478088706qp063oa.

[10] Virginia Braun and Victoria Clarke. "Can I use TA? Should I use TA? Should I *not* use TA? Comparing reflexive thematic analysis and other pattern‐based qualitative analytic approaches". In: *Counselling and Psychotherapy Research* 21.1 (Mar. 2021), pp. 37–47. ISSN: 1473-3145, 1746-1405. DOI: 10.1002/capr.12360. URL: https://onlinelibrary.wiley.com/doi/10.1002/capr.12360.

[11] Virginia Braun and Victoria Clarke. "Conceptual and design thinking for thematic analysis." In: *Qualitative Psychology* 9.1 (Feb. 2022), pp. 3–26. ISSN: 2326-3598, 2326-3601. DOI: 10.1037/qup0000196. URL: https://doi.apa.org/doi/10.1037/qup0000196.

[12] Virginia Braun and Victoria Clarke. "Toward good practice in thematic analysis: Avoiding common problems and be(com)ing a *knowing* researcher". In: *International Journal of Transgender Health* 24.1 (Jan. 25, 2023), pp. 1–6. ISSN: 2689-5269. DOI: 10.1080/26895269.2022.2129597. URL: https://www.tandfonline.com/doi/full/10.1080/26895269.2022.2129597.

[13] Hao Chen, Min Wang, and Zhen Zhang. "Research on Rural Landscape Preference Based on TikTok Short Video Content and User Comments". In: *International Journal of Environmental Research and Public Health* 19.16 (Aug. 16, 2022), p. 10115. ISSN: 1660-4601. DOI: 10.3390/ijerph191610115. URL: https://www.mdpi.com/1660-4601/19/16/10115.

[14] Camille Cobb and Tadayoshi Kohno. "How Public Is My Private Life?: Privacy in Online Dating". In: *Proceedings of the 26th International Conference on World Wide Web*. WWW '17: 26th International World Wide Web Conference. Perth Australia: International World Wide Web Conferences Steering Committee, Apr. 3, 2017, pp. 1231–1240. ISBN: 978-

[15] Jonas Colliander. "“This is fake news”: Investigating the role of conformity to other users' views when commenting on and spreading disinformation in social media". In: *Computers in Human Behavior* 97 (Aug. 2019), pp. 202–215. ISSN: 07475632. DOI: 10.1016/j.chb.2019.03.032. URL: https://linkinghub.elsevier.com/retrieve/pii/S074756321930130X.

[16] Joseph Cox ·. *The End of Privacy is a Taylor Swift Fan TikTok Account Armed with Facial Recognition Tech*. 404 Media. Sept. 25, 2023. URL: https://www.404media.co/the-end-of-privacy-is-a-taylor-swift-fan-tiktok-account-armed-with-facial-recognition-tech/.

[17] *craigslist | about | help | faqs | lifespan*. URL: https://www.craigslist.org/about/help/faqs/lifespan.

[18] Casey Fiesler. *"Participant" Perceptions of Twitter Research Ethics*. DOI: 10.1177/2056305118763366. URL: https://journals.sagepub.com/doi/epub/10.1177/2056305118763366.

[19] Casey Fiesler and Blake Hallinan. ""We Are the Product": Public Reactions to Online Data Sharing and Privacy Controversies in the Media". In: *Proceedings of the 2018 CHI Conference on Human Factors in Computing Systems*. CHI '18. New York, NY, USA: Association for Computing Machinery, Apr. 19, 2018, pp. 1–13. ISBN: 978-1-4503-5620-6. DOI: 10.1145/3173574.3173627. URL: https://dl.acm.org/doi/10.1145/3173574.3173627.

[20] Casey Fiesler, Michael Zimmer, Nicholas Proferes, Sarah Gilbert, and Naiyan Jones. "Remember the Human: A Systematic Review of Ethical Considerations in Reddit Research". In: *Proc. ACM Hum.-Comput. Interact.* 8 (GROUP Feb. 21, 2024), 5:1–5:33. DOI: 10.1145/3633070. URL: https://dl.acm.org/doi/10.1145/3633070.

[21] Linda Finlay. "Thematic Analysis: : The 'Good', the 'Bad' and the 'Ugly'". In: *European Journal for Qualitative Research in Psychotherapy* 11 (July 20, 2021), pp. 103–116. ISSN: 1756-7599. URL: https://ejqrp.org/index.php/ejqrp/article/view/136.

[22] Leah R. Fowler, Lauren Schoen, Hadley Stevens Smith, and Stephanie R. Morain. "Sex Education on TikTok: A Content Analysis of Themes". In: *Health Promotion Practice* 23.5 (Sept. 2022), pp. 739–742. ISSN: 1524-8399, 1552-6372. DOI: 10.1177/15248399211031536. URL: https://journals.sagepub.com/doi/10.1177/15248399211031536.

[23] *GeoGuessr - Let's explore the world!* URL: https://www.geoguessr.com/.

[24] Sarah Gilbert, Katie Shilton, and Jessica Vitak. "When research is the context: Cross-platform user expectations for social media data reuse". In: *Big Data & Society* (Mar. 28, 2023). Publisher: SAGE PublicationsSage UK: London, England. DOI: 10.1177/20539517231164108. URL: https://journals.sagepub.com/doi/10.1177/20539517231164108.

[25] Jennifer A. Harriger, Madeline R. Wick, Christina M. Sherline, and Abbey L. Kunz. "The body positivity movement is not all that positive on TikTok: A content analysis of body positive TikTok videos". In: *Body Image* 46 (Sept. 2023), pp. 256–264. ISSN: 17401445. DOI: 10.1016/j.bodyim.2023.06.003. URL: https://linkinghub.elsevier.com/retrieve/pii/S1740144523000852.

[26] Rakibul Hasan. "Reducing Privacy Risks in the Context of Sharing Photos Online". In: *Extended Abstracts of the 2020 CHI Conference on Human Factors in Computing Systems*. CHI EA '20. New York, NY, USA: Association for Computing Machinery, Apr. 25, 2020, pp. 1–11. ISBN: 978-1-4503-6819-3. DOI: 10.1145/3334480.3375040. URL: https://dl.acm.org/doi/10.1145/3334480.3375040.

[27] Kashmir Hill. *Two Students Created Face Recognition Glasses. It Wasn't Hard. - The New York Times*. URL: https://www.nytimes.com/2024/10/24/technology/facial-recognition-glasses-privacy-harvard.html.

[28] Roberto Hoyle, Luke Stark, Qatrunnada Ismail, David Crandall, Apu Kapadia, and Denise Anthony. "Privacy Norms and Preferences for Photos Posted Online". In: *ACM Transactions on Computer-Human Interaction* 27.4 (Aug. 31, 2020), pp. 1–27. ISSN: 1073-0516, 1557-7325. DOI: 10.1145/3380960. URL: https://dl.acm.org/doi/10.1145/3380960.

[29] Jane Im, Jill Dimond, Melody Berton, Una Lee, Katherine Mustelier, Mark S. Ackerman, and Eric Gilbert. "Yes: Affirmative Consent as a Theoretical Framework for Understanding and Imagining Social Platforms". In: *Proceedings of the 2021 CHI Conference on Human Factors in Computing Systems*. CHI '21: CHI Conference on Human Factors in Computing Systems. Yokohama Japan: ACM, May 6, 2021, pp. 1–18. ISBN: 978-1-4503-8096-6. DOI: 10.1145/3411764.3445778. URL: https://dl.acm.org/doi/10.1145/3411764.3445778.

[30] Marina C. Jenkins and Megan A. Moreno. "Vaccination Discussion among Parents on Social Media: A Content Analysis of Comments on Parenting Blogs". In: *Journal of Health Communication* 25.3 (Mar. 3, 2020), pp. 232–242. ISSN: 1081-0730, 1087-0415. DOI: 10.1080/10810730.2020.1737761. URL: https://www.tandfonline.com/doi/full/10.1080/10810730.2020.1737761.

[31] Shan Jiang and Christo Wilson. "Linguistic Signals under Misinformation and Fact-Checking: Evidence from User Comments on Social Media". In: *Proceedings of the ACM on Human-Computer Interaction* 2 (CSCW Nov. 2018), pp. 1–23. ISSN: 2573-0142. DOI: 10.1145/3274351. URL: https://dl.acm.org/doi/10.1145/3274351.

[32] Maritza Johnson, Serge Egelman, and Steven M. Bellovin. "Facebook and privacy: it's complicated". In: *Proceedings of the Eighth Symposium on Usable Privacy and Security*. SOUPS '12: Symposium On Usable Privacy and Security. Washington, D.C.: ACM, July 11, 2012, pp. 1–15. ISBN: 978-1-4503-1532-6. DOI: 10.1145/2335356.2335369. URL: https://dl.acm.org/doi/10.1145/2335356.2335369.

[33] Ji Won Kim and Gina Masullo Chen. "Exploring the Influence of Comment Tone and Content in Response to Misinformation in Social Media News". In: *Journalism Practice* 15.4 (Apr. 21, 2021), pp. 456–470. ISSN: 1751-2786, 1751-2794. DOI: 10.1080/17512786.2020.1739550. URL: https://www.tandfonline.com/doi/full/10.1080/17512786.2020.1739550.

[34] Jennifer King, Airi Lampinen, and Alex Smolen. "Privacy: is there an app for that?" In: *Proceedings of the Seventh Symposium on Usable Privacy and Security*. SOUPS '11: Symposium On Usable Privacy and Security. Pittsburgh Pennsylvania: ACM, July 20, 2011, pp. 1–20. ISBN: 978-1-4503-0911-0. DOI: 10.1145/2078827.2078843. URL: https://dl.acm.org/doi/10.1145/2078827.2078843.

[35] Daniel Klug, Yiluo Qin, Morgan Evans, and Geoff Kaufman. "Trick and Please. A Mixed-Method Study On User Assumptions About the TikTok Algorithm". In: *13th ACM Web Science Conference 2021*. WebSci '21: WebSci '21 13th ACM Web Science Conference 2021. Virtual Event United Kingdom: ACM, June 21, 2021, pp. 84–92. ISBN: 978-1-4503-8330-1. DOI: 10.1145/3447535.3462512. URL: https://dl.acm.org/doi/10.1145/3447535.3462512.

[36] Tory Lanez. *N.A.M.E.* 2015. URL: https://soundcloud.com/torylanez/03-name-prod-tory-lanez-x-play-picasso-x-happy-perez?in=apent/sets/hp.

[37] *Location information on TikTok | TikTok Help Center*. URL: https://support.tiktok.com/en/account-and-privacy/account-privacy-settings/location-services-on-tiktok#5.

[38] Steve Mann, Jason Nolan, and Barry Wellman. "Sousveillance: Inventing and Using Wearable Computing Devices for Data Collection in Surveillance Environments." In: *Surveillance & Society* 1.3 (2003), pp. 331–355. ISSN: 1477-7487. DOI: 10.24908/ss.v1i3.3344. URL: https://ojs.library.queensu.ca/index.php/surveillance-and-society/article/view/3344.

[39] Ramishah Maruf. *For Palestinian Americans and activists, doxxing is nothing new | CNN Business*. CNN. Oct. 15, 2023. URL: https://www.cnn.com/2023/10/15/business/palestinian-americans-activists-doxxing/index.html.

[40] Alice Marwick. "The Public Domain: Surveillance in Everyday Life". In: *Surveillance & Society* 9.4 (June 20, 2012), pp. 378–393. ISSN: 1477-7487. DOI: 10.24908/ss.v9i4.4342. URL: https://ojs.library.queensu.ca/index.php/surveillance-and-society/article/view/pub_dom.

[41] Alessandra Mazzia, Kristen LeFevre, and Eytan Adar. "The PViz comprehension tool for social network privacy settings". In: *Proceedings of the Eighth Symposium on Usable Privacy and Security*. SOUPS '12: Symposium On Usable Privacy and Security. Washington, D.C.: ACM, July 11, 2012, pp. 1–12. ISBN: 978-1-4503-1532-6. DOI: 10.1145/2335356.2335374. URL: https://dl.acm.org/doi/10.1145/2335356.2335374.

[42] *Missed Connections New York City | Missed You*. URL: https://missedyounyc.com/.

[43] Meera Navlakha. *TikTok is (still) obsessed with exposing cheating. But are internet sleuths going too far?* Mashable. Section: Life. June 28, 2024. URL: https://mashable.com/article/tiktok-cheating-relationships-sleuthing-viral.

[44] *new york missed connections*. craigslist. URL: https://newyork.craigslist.org/search/mis.

[45] Helen Nissenbaum. "Privacy as Contextual Integrity". In: *Washington Law Review* 79 (2004).

[46] Oladapo Oyebode, Chinenye Ndulue, Ashfaq Adib, Dinesh Mulchandani, Banuchitra Suruliraj, Fidelia Anulika Orji, Christine T Chambers, Sandra Meier, and Rita Orji. "Health, Psychosocial, and Social Issues Emanating From the COVID-19 Pandemic Based on Social Media Comments: Text Mining and Thematic Analysis Approach". In: *JMIR Medical Informatics* 9.4 (Apr. 6, 2021), e22734. ISSN: 2291-9694. DOI: 10.2196/22734. URL: https://medinform.jmir.org/2021/4/e22734.

[47] *r/NYCmissedconnections*. URL: https://www.reddit.com/r/NYCmissedconnections/.

[48] Yasmeen Rashidi, Apu Kapadia, Christena Nippert-Eng, and Norman Makoto Su. ""It's easier than causing confrontation": Sanctioning Strategies to Maintain Social Norms and Privacy on Social Media". In: *Proceedings of the ACM on Human-Computer Interaction* 4 (CSCW1 May 28, 2020), pp. 1–25. ISSN: 2573-0142. DOI: 10.1145/3392827. URL: https://dl.acm.org/doi/10.1145/3392827.

[49] Ian Rowe. "Civility 2.0: a comparative analysis of incivility in online political discussion". In: *Information, Communication & Society* 18.2 (Feb. 2015), pp. 121–138. ISSN: 1369-118X, 1468-4462. DOI: 10.1080/1369118X.2014.940365. URL: http://www.tandfonline.com/doi/abs/10.1080/1369118X.2014.940365.

[50] Joseph Scott Schafer, Brett A. Halperin, Sourojit Ghosh, and Julie Vera. "TO SCREENSHOT OR NOT TO SCREENSHOT? TENSIONS IN REPRESENTING VISUAL SOCIAL MEDIA PLATFORM POSTS". In: *AoIR Selected Papers of Internet Research* (2024). ISSN: 2162-3317. DOI: 10.5210/spir.v2024i0.14055. URL: https://spir.aoir.org/ojs/index.php/spir/article/view/14055.

[51] Evan Selinger and Hyo Joo (Judy) Rhee. "Normalizing Surveillance". In: *SATS* 22.1 (July 1, 2021). Publisher: De Gruyter, pp. 49–74. ISSN: 1869-7577. DOI: 10.1515/sats-2021-0002. URL: https://www.degruyter.com/document/doi/10.1515/sats-2021-0002/html.

[52] Daniel J. Solove. "A Taxonomy of Privacy". In: *University of Pennsylvania Law Review* 154.3 (Jan. 1, 2006), p. 477. ISSN: 00419907. DOI: 10.2307/40041279. URL: https://www.jstor.org/stable/10.2307/40041279?origin=crossref.

[53] Jessica Staddon, David Huffaker, Larkin Brown, and Aaron Sedley. "Are privacy concerns a turn-off?: engagement and privacy in social networks". In: *Proceedings of the Eighth Symposium on Usable Privacy and Security*. SOUPS '12: Symposium On Usable Privacy and Security. Washington, D.C.: ACM, July 11, 2012, pp. 1–13. ISBN: 978-1-4503-1532-6. DOI: 10.1145/2335356.2335370. URL: https://dl.acm.org/doi/10.1145/2335356.2335370.

[54] Jonathan Stempel. "Woman who called police on Black bird-watcher in Central Park loses employment appeal". In: *Reuters* (June 8, 2023). URL: https://www.reuters.com/world/us/woman-who-called-police-black-bird-watcher-central-park-loses-employment-appeal-2023-06-08/.

[55] Sophie Stephenson, Christopher Nathaniel Page, Miranda Wei, Apu Kapadia, and Franziska Roesner. "Sharenting on TikTok: Exploring Parental Sharing Behaviors and the Discourse Around Children's Online Privacy". In: *Proceedings of the CHI Conference on Human Factors in Computing Systems*. CHI '24: CHI Conference on Human Factors in Computing Systems. Honolulu HI USA: ACM, May 11, 2024, pp. 1–17. ISBN: 9798400703300. DOI: 10.1145/3613904.3642447. URL: https://dl.acm.org/doi/10.1145/3613904.3642447.

[56] Shannan Strach. "Body Image and Social Media Sharing: A Content Analysis of Public Reactions to a Body Positive Post on TikTok". In: *Crossing Borders: Student Reflections on Global Social Issues* 5.1 (May 24, 2023). Number: 1. ISSN: 2562-8445. DOI: 10.31542/cb.v5i1.2528. URL: https://journals.macewan.ca/crossingborders/article/view/2528.

[57] Kurt Thomas, Devdatta Akhawe, Michael Bailey, Dan Boneh, Elie Bursztein, Sunny Consolvo, Nicola Dell, Zakir Durumeric, Patrick Gage Kelley, Deepak Kumar, Damon McCoy, Sarah Meiklejohn, Thomas Ristenpart, and Gianluca Stringhini. "SoK: Hate, Harassment, and the Changing Landscape of Online Abuse". In: *2021 IEEE Symposium on Security and Privacy (SP)*. 2021 IEEE Symposium on Security and Privacy (SP). ISSN: 2375-1207. May 2021, pp. 247–267. DOI: 10.1109/SP40001.2021.00028. URL: https://ieeexplore.ieee.org/document/9519435/?arnumber=9519435.

[58] Stephanie Thompson. *Creator of Ohio 'Peach Bowl Girl' TikTok shares message for her*. URL: https://www.nbc4i.com/news/local-news/ohio-state-university/creator-of-ohio-peach-bowl-girl-tiktok-shares-message-for-her/.

[59] *TikTok Community Guidelines, Privacy and Security*. URL: https://www.tiktok.com/community-guidelines/en/privacy-security.

[60] Samuel D. Warren and Louis D. Brandeis. "The Right to Privacy". In: *Harvard Law Review* 4.5 (Dec. 15, 1890), p. 193. ISSN: 0017811X. DOI: 10.2307/1321160. URL: https://www.jstor.org/stable/1321160?origin=crossref.

[61] Miranda Wei, Eric Zeng, Tadayoshi Kohno, and Franziska Roesner. "Anti-privacy and anti-security advice on tiktok: case studies of technology-enabled surveillance and control in intimate partner and parent-child relationships". In: *Proceedings of the Eighteenth USENIX Conference on Usable Privacy and Security*. SOUPS'22. USA: USENIX Association, Aug. 8, 2022, pp. 447–462. ISBN: 978-1-939133-30-4.

[62] Jason Whiting, Rachael Dansby Olufowote, Jaclyn Cravens-Pickens, and Alyssa Banford Witting. "Online Blaming and Intimate Partner Violence: A Content Analysis of Social Media Comments". In: *The Qualitative Report* (Jan. 13, 2019). ISSN: 2160-3715, 1052-0147. DOI: 10.46743/2160-3715/2019.3486. URL: https://nsuworks.nova.edu/tqr/vol24/iss1/6/.

[63] Yuhao Yang, Jonathan Lutes, Fengjun Li, Bo Luo, and Peng Liu. "Stalking online: on user privacy in social networks". In: *Proceedings of the second ACM conference on Data and Application Security and Privacy*. CODASPY'12: Second ACM Conference on Data and Application Security and Privacy. San Antonio Texas USA: ACM, Feb. 7, 2012, pp. 37–48. ISBN: 978-1-4503-1091-8. DOI: 10.1145/2133601.2133607. URL: https://dl.acm.org/doi/10.1145/2133601.2133607.

[64] Michael Zimmer. ""But the data is already public": on the ethics of research in Facebook". In: *Ethics and Information Technology* 12.4 (Dec. 1, 2010), pp. 313–325. ISSN: 1572-8439. DOI: 10.1007/s10676-010-9227-5. URL: https://doi.org/10.1007/s10676-010-9227-5.

[65] Diana Zulli and David James Zulli. "Extending the Internet meme: Conceptualizing technological mimesis and imitation publics on the TikTok platform". In: *New Media & Society* 24.8 (Aug. 2022), pp. 1872–1890. ISSN: 1461-4448, 1461-7315. DOI: 10.1177/1461444820983603. URL: https://journals.sagepub.com/doi/10.1177/1461444820983603.

# Appendix

Table 4: Our codebook for comment analysis. *Other reactions were excluded in analysis and are not presented in Section 5.2.

| Theme | Sub-theme | # | Code | # | Code Description | Example |
|---|---|---|---|---|---|---|
| Supportive reactions (n=883) | Search party | 235 | Identification | 186 | (Attempts to) identify the subject | "that's my cousin john" |
| | | | Advice | 49 | Giving the creator advice on how to find or interact with the subject; requesting additional information | "which hotel was he seen at?" |
| | Communal support | 217 | Encouragement | 181 | Offering words of encouragement; supporting the search | "this is so cute, I hope you find him!" |
| | | | TikTok platform | 36 | Referencing the TikTok platform with respect to the search | "TikTok, do your magic!" |
| | Algorithmic engagement | 175 | Location | 124 | Referencing the location of the commenter (to demonstrate audience location) | "You've reached Atlanta!" |
| | | | Boosting | 51 | Commenting to increase engagement and thus further algorithmic recommendations of the video; using other engagement tactics | "commenting to boost" |
| | Suspense | 166 | Request for update | 84 | Expressing interest in an update on the search | "update?! did you find her?" |
| | | | Investment | 82 | Indicating investment in the search and/or romantic outcomes | "did you meet up?????!" |
| | Relatability | 166 | Empathy | 98 | Expressing that the commenter relates or empathizes with the video | "this is so me.. I'm way too awkward to just walk up to someone and say hi" |
| | | | Subject attractiveness | 68 | Expressing that the subject is attractive | "idk who that is, but he's so cute!" |
| Disapproving reactions (n=310) | Anti-social norms | 144 | Proximity | 74 | Expressing how the subject was in close proximity to the creator | "hey so you're actually right there hope this helps" |
| | | | Interaction | 52 | Expressing that the creator should have spoken to the subject instead of posting the video | "why didn't you just talk to her..." |
| | | | Social norms | 18 | Referencing social norms or commenting on perceived (ab)normal behavior | "Do y'all think going up to them is more awkward then finding them by posting them on tiktok..." |
| | Relationship | 89 | Relationship status | 46 | Referencing the (imagined) relationship status of the subject | "imagine he has a girlfriend already..." |
| | | | Age | 25 | Referencing the age of the subject, especially in relation to the creator | "aren't they a little young?" |
| | | | Relationship advice | 18 | Giving discouraging relationship advice regarding the subject, especially from commenter personal experience | "From personal experience... run girl" |
| | Fear factor | 76 | Negative judgments | 40 | Expressing fear, concern, or other negative judgments and other discouraging language about the search | "It's concerning how you all know so much about him" |
| | | | Stalking | 26 | Referencing stalkers or the act of stalking | "did u really just stalk him…" |
| | | | Surveillance | 10 | Referencing surveillance or surveillance professions, e.g., police, FBI | "TIKTOK COMMENTS ARE BETTER THAN THE POLICE" |
| | Gender | 17 | Gender | 17 | Referencing gender norms or gendered double standards | "if a man did this, they're going to jail." |
| | Consent | 16 | Consent | 16 | Referencing consent, permission, or respect | "please don't take videos of ppl without their consent" |
| Other* (n=708) | Ambiguous | 691 | - | - | - | - |
| | Mixed | 17 | - | - | - | - |